# Effect of electron- and hole-doping on properties of kagomé-lattice ferromagnet $Fe_3Sn_2$

Milo Adams[1], Chen Huang[2,3,4,*] and Michael Shatruk[1,3,4,*]

[1] Department of Chemistry and Biochemistry, Florida State University, Tallahassee, FL 32306-4390, USA
[2] Department of Scientific Computing, Florida State University, Tallahassee, FL 32306-4120, USA
[3] Materials Science and Engineering Program, Florida State University, Tallahassee, FL 32310, USA
[4] National High Magnetic Field Laboratory, Tallahassee, FL 32310, USA

E-mail: shatruk@chem.fsu.edu and chuang3@fsu.edu

## Abstract

We report a theoretical investigation of effects of Mn and Co substitution in the transition metal sites of the kagomé-lattice ferromagnet, $Fe_3Sn_2$. Herein, hole- and electron-doping effects of $Fe_3Sn_2$ have been studied by density-functional theory calculations on the parent phase and on the substituted structural models of $Fe_{3-x}M_xSn_2$ (M = Mn, Co; $x$ = 0.5, 1.0). All optimized structures favor the ferromagnetic ground state. Analysis of the electronic density of states (DOS) and band structure plots reveals that the hole (electron) doping leads to a progressive decrease (increase) in the magnetic moment per Fe atom and per unit cell overall. The high DOS is retained nearby the Fermi level in the case of both Mn and Co substitutions. The electron doping with Co results in the loss of nodal band degeneracies, while in the case of hole doping with Mn emergent nodal band degeneracies and flatbands initially are suppressed in $Fe_{2.5}Mn_{0.5}Sn_2$ but re-emerge in $Fe_2MnSn_2$. These results provide key insights into potential modifications of intriguing coupling between electronic and spin degrees of freedom observed in $Fe_3Sn_2$.

Supplementary material for this article is available online

(Some figures may appear in colour only in the online journal)

## 1. Introduction

The kagomé lattice has served as one of the main testbeds for studies on the physics of spin frustration [1,2], allowing fundamental theoretical and experimental insights into such unconventional states of matter as spin ice [3] and spin liquid [3,4]. Pairwise antiferromagnetic (AFM) interactions between nearest neighbors cannot be simultaneously satisfied in the kagomé-lattice arrangement of magnetic moments [5], as AFM alignment of two spins on a single triangle leads to ambiguous orientation of the third spin on the same triangle, thus disrupting magnetic ordering [6]. Such frustration propagating through a solid-state lattice leads to fragile magnetism, i.e., even small external perturbations might have a strong effect on the long-range magnetic behavior across the periodic structure [7,8]. Thus, the kagomé-type materials hold great potential not only for insight into fundamental properties of spin-frustrated systems but also for practical applications in future spintronic devices with low-energy power consumption [9].

The recent growth of research on topological materials has reinvigorated interest to the kagomé lattice due to its tendency to host unique electronic correlations [6,10]. The network of hexagons enclosed by corner-sharing triangles creates the topological condition for destructive interference of the electron wavefunctions around each hexagon. As a result, dispersionless flatbands are observed in the electronic band structure [11-14]. Charge carriers occupying these highly degenerate flatbands are rendered "superheavy" by self-localization. Driven by these considerations, research efforts on kagomé lattices have ventured beyond AFM-correlated materials, as even ferromagnetic (FM) kagomé materials have been shown to exhibit intriguing topological properties [15]. For example, a giant anomalous Hall effect [16], massive Dirac fermions [17], and skyrmion bubbles [18] have been reported for the kagomé-lattice metal, $Fe_3Sn_2$. This material crystallizes in the filled NiAs structure type [19] (the space group $R\bar{3}m$). The crystal structure features kagomé layers of Fe atoms alternating with hexagonal layers of Sn atoms, while additional Sn atoms center hexagons of the kagomé layer (Figure 1). A 2009 report classifying $Fe_3Sn_2$ as a frustrated ferromagnet with the ordering temperature ($T_C$) of 640 K and temperature-dependent non-collinear spin texture [20], followed by the report of the giant anomalous Hall effect [16], spurred extensive studies of the electronic structure and topological properties of this material. Dirac cones at symmetry point K were revealed experimentally by ARPES measurements [17], while further analysis identified dispersionless flatbands at ~0.2 eV below the Fermi level along the Γ-K and Γ-M directions [21]. Isolation of these bands is of great interest for the realization of new magnetically driven fractionalized phases of matter. A possible approach to achieve this objective is to tune the electronic structure by chemical substitutions. Till now, however, essentially all studies of $Fe_3Sn_2$ have focused on the pristine composition.

In this work, we report a theoretical evaluation of the effects of hole and electron doping on the magnetic properties and electronic structure of $Fe_3Sn_2$. These effects are achieved by partial replacement of Fe with Mn and Co, respectively. The compositional analogues, $Mn_3Sn_2$ [22] and $Co_3Sn_2$ [23], are known ferromagnets that belong to the orthorhombic $Ni_3Sn_2$ structure type (space group *Pnma*), although the high-temperature modification of $Co_3Sn_2$, obtained by quenching from above 900 K, does exhibit the hexagonal filled NiAs-type structure [23]. Effects of Fe for Mn substitution on the properties of $Mn_3Sn_2$ were studied by Recour *et al.* in a report on magnetic and magnetocaloric properties of $Fe_{3-x}M_xSn_2$ on the Mn-rich side (0.1 < *x* < 0.8) [24], but no such studies have been performed for the other limiting composition, i.e., when substituting Mn for Fe in $Fe_3Sn_2$. The present work, to the best of our knowledge, is the first attempt to explore possible effects of chemical substitutions on the properties of $Fe_3Sn_2$.

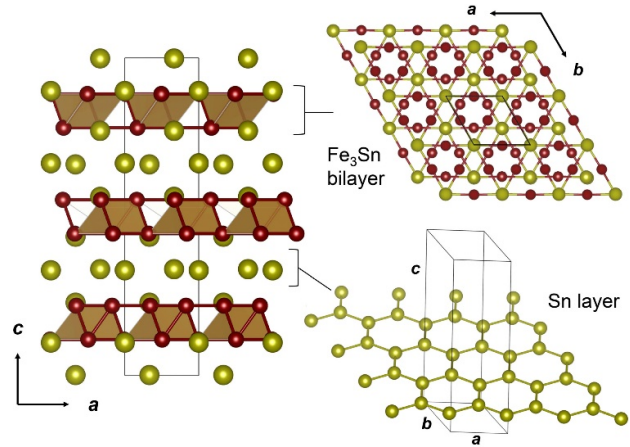

**Figure 1.** The crystal structure of $Fe_3Sn_2$ viewed perpendicular to the *ac* plane (left) to reveal alternating $Fe_3Sn$ and Sn layers, shown on the right. The $Fe_3Sn$ slab is formed by stacking of two kagomé layers of Fe atoms, with the Sn atoms centering hexagons of each layer. In the image of the bilayer, the brighter and duller colors indicate atoms located in different kagomé layers.

## 2. Methods

Electronic structure calculations were performed at the density-functional theory (DFT) level with Quantum Espresso (version 7.0) [25], using scalar relativistic projector-augmented wave pseudopotentials within the Perdew-Burke-Ernzerhof generalized gradient approximation (PBE-GGA) [26]. The self-consistent energy convergence criterion was set to $10^{-6}$ eV, with a Γ-centered 8x8x8 *k*-point mesh. The reported unit cell and atomic parameters of $Fe_3Sn_2$ were taken as the starting point for calculations [19]. Lattice relaxations were conducted with convergence thresholds of 0.5 kbar for pressure, 0.001 eV/Å for forces, and $10^{-5}$ eV for the total energy per unit cell. Fermi-Dirac smearing was used with a smearing energy of 0.1 eV. The plane wave kinetic energy cutoff was set at 500 eV. A dense 16x16x16 *k*-point mesh was used to calculate the density of states. Crystal structure visualization and plotting were performed with VESTA [27].

The influence of hole- and electron-doping effects on the electronic structure of $Fe_3Sn_2$ was modeled by partially substituting Mn and Co, respectively, for Fe. Since we are interested in perturbations of the electronic structure in the vicinity of the Fermi level, and since $M_3Sn_2$ (M = Mn, Co) exhibit a different structure type, we have restricted these calculations to the smaller levels of substitution, i.e., to compositions $Fe_{3-x}M_xSn_2$ with *x* = 0.5 and 1.0. To implement such models, the 30 atoms in the unit cell of $Fe_3Sn_2$ were described as unique positions by lowering the symmetry to *P*1. Calculations performed on $Fe_3Sn_2$ in both the original $R\bar{3}m$ and the *P*1 space groups yielded comparable results, thus confirming the validity of such an approach.

## 2. Results and Discussion

### 2.1 Substitution models

Our initial DFT calculations on the unsubstituted $Fe_3Sn_2$ revealed that the FM state is stabilized by 0.089 eV per unit cell relative to the paramagnetic (PM) state. The calculated magnetic moments were 2.24 $\mu_B$ per Fe atom and –0.08 $\mu_B$ per Sn atom, giving the total magnetic moment of 6.56 $\mu_B$ per formula unit (f.u.). These values are essentially the same as those reported recently by Fayyazi *et al.* [28]. They also compare well to the magnetic moment of 2.17 $\mu_B$ per Fe atom determined by neutron diffraction experiments at 10 K [29]. The most pronounced features of the band structure near the Fermi level are also commensurate with the previous reports, as will be discussed below in the section on the hole- and electron-doping effects.

Given the good agreement between our calculations and previous experimental and theoretical results on $Fe_3Sn_2$, we proceeded to calculating the electronic structures and magnetic properties of $Fe_{3-x}M_xSn_2$ (M = Mn, Co; $x$ = 0.5 and 1.0). To model such substitutions, we lowered the symmetry to $P1$ and confirmed that the electronic structure and magnetic parameters of $Fe_3Sn_2$ remained essentially the same when calculated in this space group. Then, the substitution models were calculated for several random distributions of the Fe and M atoms, according to the desired compositions. The similar energies of different configurations, with the maximum energy difference of 7.1 µeV, confirmed that neither Mn nor Co atoms showed preference for a specific substitution pattern (Table S1). Subsequent calculations were performed using a single substitution pattern for each of the studied compositions; the pattern was chosen in such a way as to achieve the most random distribution of the Fe and M atoms (Figure S1 and Tables S2 and S3).

The optimized crystal structures showed that the replacement of Fe with Mn or Co led, respectively, to the increase and decrease in the unit cell volume (Table 1), in agreement with expectations based on the relative size of the transition metal atoms. The only exception is the small increase in the unit cell volume when going from $Fe_3Sn_2$ to $Fe_{2.5}Co_{0.5}Sn_2$. Interestingly, the hole- and electron-doping have different effects on the average distances between the transition metal atoms. The experimental $Fe_3Sn_2$ structure has two types of such distances: there are shorter (2.590 Å) and longer (2.754 Å) Fe–Fe distances within the kagomé layer and the same shorter distance (2.590 Å) between the layers in the bilayer kagomé structure. Since we performed the crystal structure optimization without symmetry restrictions, we observe three different values for two shorter and one longer distances (Table 1). The calculated shorter *intra*layer distance shows a pronounced increase, from 2.551 Å to 2.623 Å, upon substitution of Mn for Fe but is less affected by the substitution of Co for Fe, varying between 2.537 Å and 2.575 Å. On the contrary, the shorter *inter*layer distance shows smaller changes, varying between 2.556 Å and 2.587 Å upon Mn doping and between 2.556 Å and 2.549 Å upon Co doping. Although the largest changes in the distances between transition metals appear to take place within the kagomé layer, the unit cell shows much larger changes along the $c$ axis, i.e., perpendicular to the layers. This difference is explained by the pronounced changes in the distances between the transition metal and tin atoms along the $c$ axis (Table S4).

**Table 1.** Unit cell parameters and interatomic distances in the optimized structures of $Fe_{3-x}M_xSn_2$ (M = Mn, Co; $x$ = 0.5, 1.0).

| Composition | Unit cell params. | | $d$(M–M) (Å)[a] | |
|---|---|---|---|---|
| | $a, c$ (Å) | $V$ (Å$^3$) | *Intra*layer | *Inter*layer |
| $Fe_2MnSn_2$ | 5.353 19.836 | 492.24 | 2.623, 2.731 | 2.587 |
| $Fe_{2.5}Mn_{0.5}Sn_2$ | 5.338 19.835 | 489.46 | 2.607, 2.777 | 2.581 |
| $Fe_3Sn_2$ (calc.) | 5.327 19.778 | 486.04 | 2.551, 2.777 | 2.556 |
| (*exp.*) | (5.344 19.845) | (490.81) | (2.590, 2.754) | (2.590) |
| $Fe_{2.5}Co_{0.5}Sn_2$ | 5.338 19.732 | 486.92 | 2.537, 2.791 | 2.549 |
| $Fe_2CoSn_2$ | 5.320 19.556 | 479.33 | 2.575, 2.745 | 2.552 |

[a] The average distances between transition metal atoms.

### 2.2 Magnetic properties of $Fe_{3-x}M_xSn_2$

Results of calculations on the effects of hole and electron doping on the electronic structure of $Fe_3Sn_2$ are summarized in Table 2. The FM ground state remains favorable for all substituted compositions, as can be seen from the values of $\Delta E_{FM-PM}$, defined as the difference in the calculated total energy of the spin-polarized (FM) and non-spin-polarized (PM) models (the negative value of this difference indicates the stabilization provided by the FM configuration). The relative stability of the FM state decreases only slightly upon substitution of Mn for Fe, but the substitution of Co for Fe causes a substantial decrease in the magnitude of $|\Delta E_{FM-PM}|$.

Upon substitution of Mn for Fe (the hole doping), the magnetic moment per Mn atom, $m$(Mn), increases from 2.35 $\mu_B$ in $Fe_{2.5}Mn_{0.5}Sn_2$ to 2.37 $\mu_B$ in $Fe_2MnSn_2$. Meanwhile, the magnetic moment per Fe atom, $m$(Fe), initially decreases from 2.24 $\mu_B$ in $Fe_3Sn_2$ to 2.16 in $Fe_{2.5}Mn_{0.5}Sn_2$, and then increases to 2.19 in $Fe_2MnSn_2$. This increase in the magnetic moment per Fe atom leads to anomalous behavior of the total moment per f.u., which initially decreases from 6.56 $\mu_B$ in $Fe_3Sn_2$ to 6.42 $\mu_B$ in $Fe_{2.5}Mn_{0.5}Sn_2$ but then increases to 6.59 $\mu_B$ in $Fe_2MnSn_2$. Magnetic properties of Co-substituted (electron-doped) $Fe_3Sn_2$ exhibit an opposite trend, as $m$(Fe) increases and $m$(Co) decreases with the increasing Co content. As a result, the total moment per f.u. decreases from 6.03 $\mu_B$ in $Fe_{2.5}Co_{0.5}Sn_2$ to 5.21 $\mu_B$ in $Fe_2CoSn_2$.

**Table 2.** Calculated magnetic properties of $Fe_{3-x}M_xSn_2$ (M = Mn, Co; $x$ = 0.5 and 1.0).

| Composition ⇒ <br> Calcd. parameter | $Fe_2MnSn_2$ | $Fe_{2.5}Mn_{0.5}Sn_2$ | $Fe_3Sn_2$ | $Fe_{2.5}Co_{0.5}Sn_2$ | $Fe_2CoSn_2$ |
|---|---|---|---|---|---|
| $\Delta E_{FM-PM}$, meV/cell | −0.089 | −0.093 | −0.099 | −0.085 | −0.074 |
| $m(M)$, $\mu_B$ | 2.37 | 2.35 | – | 1.04 | 0.81 |
| $m(Fe)$, $\mu_B$ | 2.19 | 2.16 | 2.24 | 2.25 | 2.26 |
| $m(Sn)$, $\mu_B$ | −0.08 | −0.08 | −0.08 | −0.06 | −0.06 |
| $m(f.u.)$, $\mu_B$ | 6.59 | 6.42 | 6.56 | 6.03 | 5.21 |

Examination of the majority- and minority-spin partial DOS (pDOS) curves of the 3$d$ electrons near the Fermi level provides a justification for these observations. In the unsubstituted $Fe_3Sn_2$, the spin polarization leads to the Fermi level falling in a pseudo-gap between the nearly completely populated majority-spin states (shown with red curves in Figure 2) and the minority-spin states (shown with blue curves), a large fraction of which remains depopulated. Such a scenario justifies the stabilization of the FM ground state in this compound. The pDOS curves near the Fermi level exhibit obvious changes in the population of the minority-spin states as the value of $x$ is increased. Substitution of Mn for Fe destabilizes the minority-spin states in $Fe_{2.5}Mn_{0.5}Sn_2$ leading to an increase in moment per f.u. The minority-spin states are further destabilized in $Fe_2MnSn_2$ while the majority-spin states remain relatively unchanged, further increasing the moment per f.u. (Figures 2a-b). On the other hand, substitution of Co gradually fills the minority states at $E_F$ contributing to a consistently lower magnetic moment per f.u. with increasing Co content (Figures 2d-e). High-density pDOS peaks indicate localization of electrons leading to high effective masses and low mobility; $Fe_2MnSn_2$ exhibits the greatest energy dispersion of states, suggesting it may possess the most favorable conductive properties. Indeed, these expectations are confirmed by examination of features in the electronic band structures, which we discuss next.

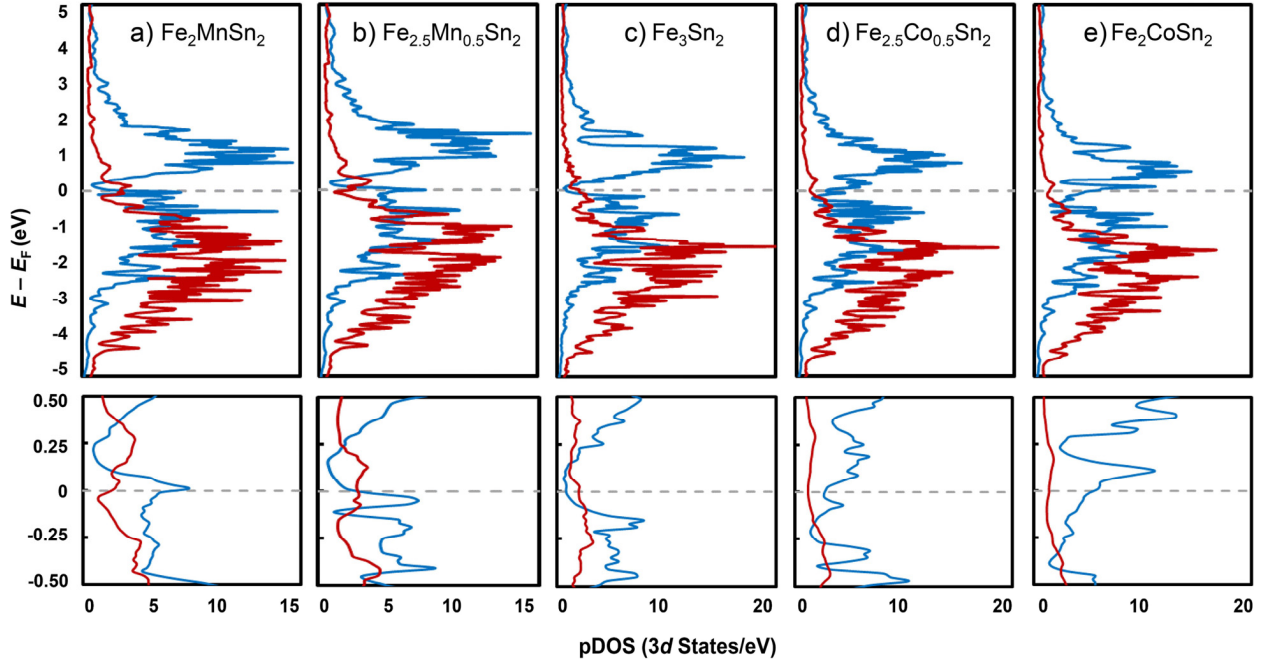

**Figure 2.** Partial DOS of the majority-spin (red) and minority-spin (blue) 3$d$ electrons of the transition metal sites in $Fe_{3-x}M_xSn_2$ (the Fermi level is indicated with a dashed gray line). Expanded plots of the region near $E_F$ are shown in the bottom panel.

*2.3 Hole- and electron-doping effects*

As already mentioned above, the features observed in the calculated electronic band structure of $Fe_3Sn_2$ are comparable to those reported in previous theoretical studies of this compound. Thus, a set of Weyl nodes centered around the symmetry point Γ is located at ~0.4 eV below $E_F$, and a set of Dirac points is observed along the V–Γ–T $k$-path at ~0.1 eV below $E_F$ (Figure 3a). These degeneracies are accompanied by high average dispersion of energy bands and large hole pockets at Γ.

The substitution of Mn for Fe has remarkable effects on the electronic band structure. In the case of $Fe_{2.5}Mn_{0.5}Sn_2$, there is a loss of the Dirac points that were observed in $Fe_3Sn_2$ at ~0.1 eV below $E_F$ along the Γ–V and Γ–T lines and an opening gap of ~0.02 eV between the Weyl nodes at ~0.4 eV below $E_F$ (indicated with red ovals in Figure 3b). A low-dispersion flatband, however, is observed where these nodes previously resided. Moreover, additional flatbands emerge across the *k*-path at the same energy, and bands below these points exhibit lower average dispersion, consistent with the loss of symmetry and increased localization of the electron wavefunctions.

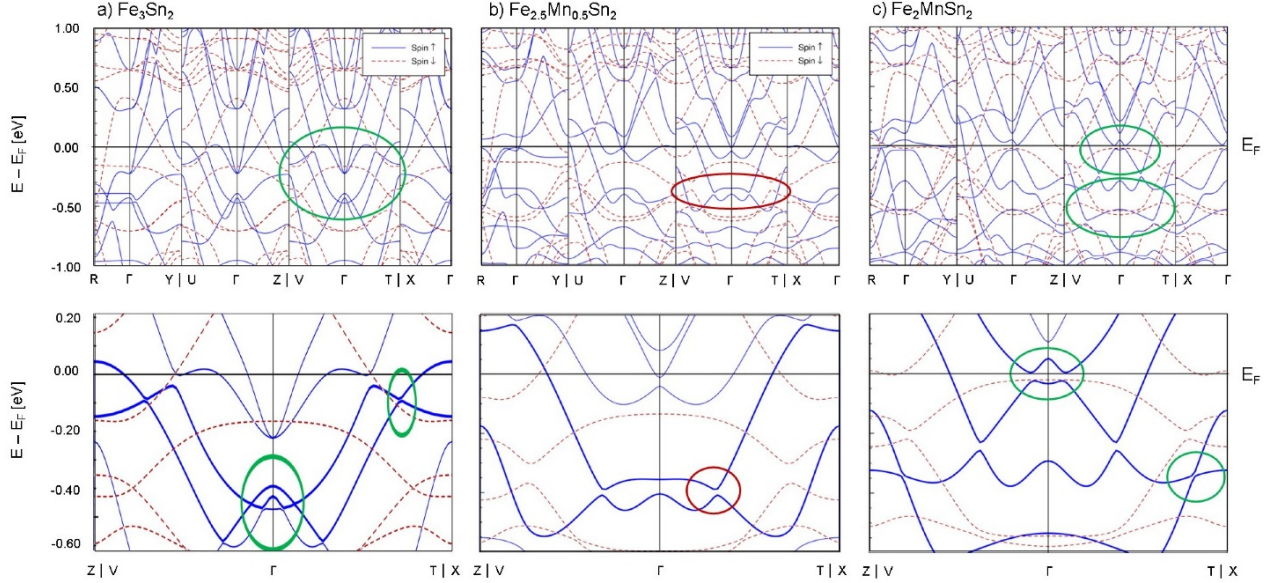

**Figure 3.** Calculated electronic band structures of $Fe_{3-x}Mn_xSn_2$, with enlarged areas near the Fermi level shown in the bottom panel. Degeneracies observed in $Fe_3Sn_2$ (a) are broken in $Fe_{2.5}Mn_{0.5}Sn_2$ (b) but appear to re-emerge in $Fe_2MnSn_2$ (c) alongside the emergence of new nodal and linear degeneracies at and below $E_F$. The solid and dashed curves correspond to the majority-spin and minority-spin states, respectively. The green ovals highlight the desirable degeneracies while the red ones indicate the potential loss of those degeneracies.

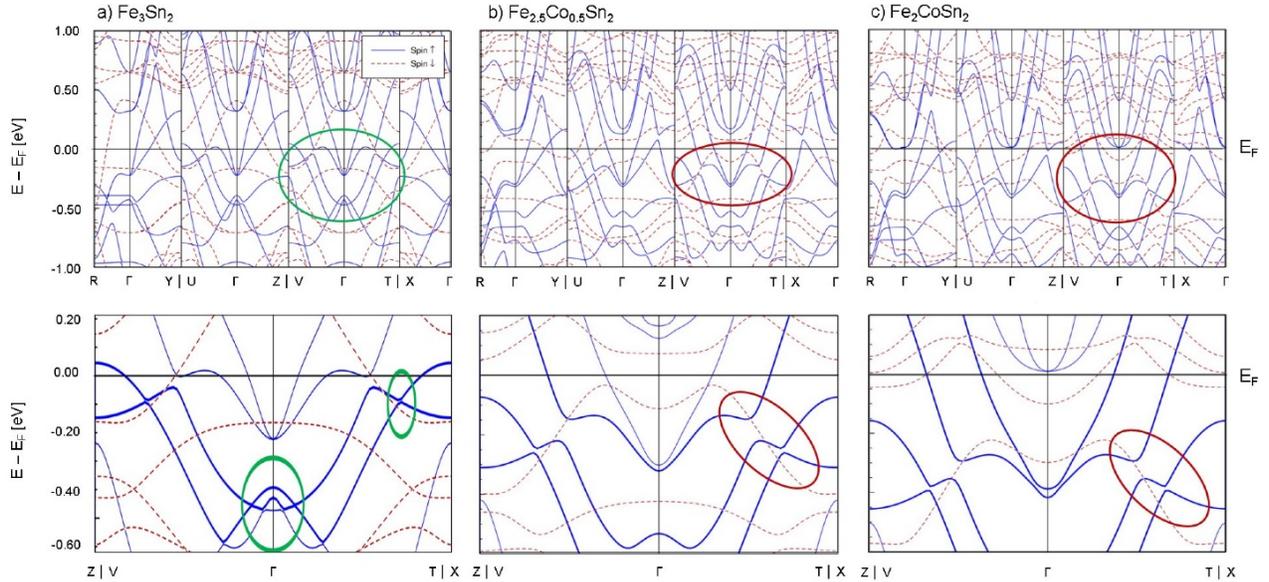

**Figure 4.** Calculated electronic band structures of $Fe_{3-x}Co_xSn_2$, with enlarged areas near the Fermi level shown in the bottom panel. Degeneracies observed in $Fe_3Sn_2$ (a) are split in $Fe_{2.5}Co_{0.5}Sn_2$ (b) and remain non-degenerate in $Fe_2CoSn_2$ (c). The solid and dashed curves correspond to the majority-spin and minority-spin states, respectively. The green ovals highlight the desirable degeneracies while the red ones indicate the potential loss of those degeneracies.

Upon further increase in the Mn content, i.e., in $Fe_2MnSn_2$, the degeneracies originally observed in $Fe_3Sn_2$ and broken in $Fe_{2.5}Mn_{0.5}Sn_2$ appear to re-emerge as potential Weyl-type and Dirac-type points apparent at $E_F$ and ~0.3 eV below $E_F$, respectively (indicated with green ovals in Figure 3c). The emergent Weyl-type cones at $E_F$ are separated by a gap of ~0.03 eV, and the gap between the former Dirac points at ~0.4 eV below $E_F$ also opens to ~0.03 eV. Nevertheless, a new set of Dirac points emerges approximately 0.05 eV below on either side of these split nodes along the V–Γ–T $k$-path. The appearance of degenerate states and high dispersion in the electronic band structure, most likely, correlate with the more symmetric atomic arrangement in the unit cell of $Fe_2MnSn_2$ as compared to that of $Fe_{2.5}Mn_{0.5}Sn_2$ (Figure S1). These changes result in greater electron delocalization which promotes increased interaction through FM exchange. Delocalization also facilitates the emergence of degenerate electronic states with potential for high carrier mobility. The expected retention of FM ordering and the high density of degenerate states near $E_F$ make $Fe_2MnSn_2$ a promising candidate for further studies, suggesting that hole-doping of $Fe_3Sn_2$, in general, may yield non-trivial physical properties.

Electronic band structures of $Fe_{3-x}Co_xSn_2$ are shown in Figure 4, where the band structure of $Fe_3Sn_2$ is also reproduced for the sake of comparison (Figure 4a). As in the case of $Fe_{2.5}Mn_{0.5}Sn_2$, the Dirac points observed for $Fe_3Sn_2$ are lost in $Fe_{2.5}Co_{0.5}Sn_2$ (Figure 4b), and two sets of cones are open with gaps of 0.01 and 0.07 eV (indicated with red ovals). Low-dispersion bands are observed across the $k$-path at ~0.7 eV below $E_F$. Further substitution of Co for Fe maintains these energy gaps between the previously degenerate points along the Γ–V and Γ–T lines, as can be seen in the electronic structure of $Fe_2CoSn_2$ (Figure 4c), although $E_F$ is shifted to a higher energy relative to these points, due to filling of the $3d$ states. Thus, the band structure of $Fe_2CoSn_2$ does not exhibit a recovery of emergent degeneracies, as was observed for $Fe_2MnSn_2$. In fact, the 0.01 eV and 0.07 eV gaps between the cones observed in $Fe_{2.5}Co_{0.5}Sn_2$ are nearly unaffected in the band structure of $Fe_2MnSn_2$. While nodal degeneracies appear at ~0.9 eV below $E_F$, such degeneracies between valence bands are relatively trivial and consistent with the more symmetric atomic arrangement. For both $Fe_{2.5}Co_{0.5}Sn_2$ and $Fe_2CoSn_2$, the lower average dispersion near $E_F$, as compared to that in $Fe_3Sn_2$, indicates reduced carrier mobility with higher effective masses. These results suggest that electron doping does not enhance the magnetic or conductive properties of $Fe_3Sn_2$, but rather suppresses these characteristics and removes non-trivial features of the band structure.

## 3. Concluding Remarks

Our theoretical investigation clearly reveals the immediate impact of hole- and electron-doping substitutions on the magnetic and electronic properties of frustrated kagomé ferromagnet $Fe_3Sn_2$. Hole-doping through substitution of Mn for Fe initially breaks the nodal degeneracies observed in $Fe_3Sn_2$, but at higher Mn content such degeneracies re-emerge. Especially interesting is the appearance of nearly degenerate Weyl points at the Fermi level, which could lead to exotic conducting and magnetic properties. The retention and emergence of degenerate states might indicate strong coupling between the electronic and spin degrees of freedom in the kagomé lattice, which could result both in anomalous transport effects and multiple magnetic ground states with non-collinear spin textures. In contrast, electron doping eliminates the nodal degeneracies and does not lead to appearance of any re-emergent features of interest in the electronic band structure. Thus, we conclude that the present study identifies $Fe_{3-x}Mn_xSn_2$ as a suitable candidate for further investigations of magnetic and topological properties of $Fe_3Sn_2$-based kagomé metals. Experimental studies of such hole-doping effects are currently under way in our laboratories, and their results will be reported in due course.

## Acknowledgements

This work was supported by the National Science Foundation (award DMR-1905499 to M.S.). The computing for this project was performed on the HPC cluster at the Research Computing Center at the Florida State University.

## Conflict of Interest Statement

There are no conflicts of interest to declare.

## Data availability statement

All computational data discussed in this work are available freely at https://github.com/shatruk-fsu/Fe3-xMxSn2.

# Supporting Information

# Effect of electron- and hole-doping on properties of kagomé-lattice ferromagnet Fe$_3$Sn$_2$

Milo Adams[1], Chen Huang[2,3,4,]* and Michael Shatruk[1,3,4,]*

[1] Department of Chemistry and Biochemistry, Florida State University, Tallahassee, FL 32306-4390, USA
[2] Department of Scientific Computing, Florida State University, Tallahassee, FL 32306-4120, USA
[3] Materials Science and Engineering Program, Florida State University, Tallahassee, FL 32310, USA
[4] National High Magnetic Field Laboratory, Tallahassee, FL 32310, USA

E-mail: shatruk@chem.fsu.edu and chuang3@fsu.edu

**Table S1.** Energies of different atomic configurations for Fe$_{2.5}$M$_{0.5}$Sn$_2$ (M = Mn, Co) relative to the ground state at 0 eV as a function of substituting the Mn or Co atom for different crystallographic sites of Fe in the primitive unit cell.

| Site[a] | | $E_{total}$ (meV) | |
|---|---|---|---|
| | | Mn | Co |
| 1 | $-y, x-y, z$ | 0.0071 | 0.0041 |
| 2 | $-x, -y, -z$ | 0.0000 | 0.0036 |
| 3 | $x, y, z$ | 0.0005 | 0.0019 |
| 4 | $-x+y, -x, z$ | 0.0007 | 0.0031 |
| 5 | $x-y, x, -z$ | 0.0033 | 0.0022 |
| 6 | $y, -x+y, -z$ | 0.0027 | 0.0000 |

[a] The coordinate labels of six sites correspond to the symmetry-related positions of Fe atoms in the original hexagonal unit cell of Fe$_3$Sn$_2$.

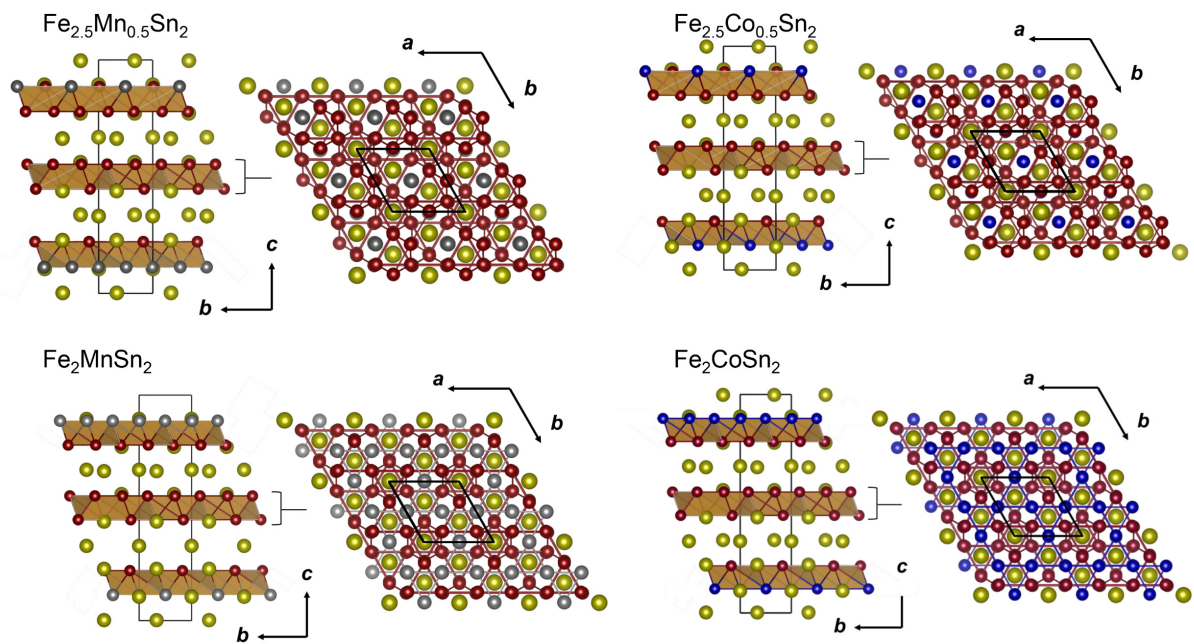

**Figure S1.** Relaxed structures of $Fe_{3-x}M_xSn_2$ viewed down the $a$ axis (left) to show the stacking of layers, and down the $c$ axis (right) to show the top view of the kagomé layer. Color scheme: Fe = garnet, Sn = gold, Mn = silver, Co = blue.

**Table S2.** An example configuration of Fe and Mn atoms in the relaxed $Fe_{3-x}Mn_xSn_2$ unit cell for Mn substitution in Site 2 (Table S1), as determined by DFT calculations. The positions are given as fractional coordinates.

| $Fe_{2.5}Mn_{0.5}Sn_2$ | | | | $Fe_2MnSn_2$ | | | |
|---|---|---|---|---|---|---|---|
| Atom | x | y | z | Atom | x | y | z |
| Mn1 | 0.4972 | 0.5071 | 0.1130 | Mn1 | 0.4966 | 0.5034 | 0.1136 |
| Mn2 | 0.5074 | 0.4962 | 0.8866 | Mn2 | 0.5034 | 0.4967 | 0.8863 |
| Mn3 | 0.4968 | 0.9913 | 0.1126 | Mn3 | 0.4966 | 0.9933 | 0.1136 |
| Fe1 | 0.5105 | 0.0166 | 0.8860 | Mn4 | 0.5034 | 0.0067 | 0.8863 |
| Fe2 | 0.0167 | 0.5093 | 0.1133 | Mn5 | 0.0067 | 0.5034 | 0.1136 |
| Fe3 | 0.9868 | 0.4930 | 0.8865 | Mn6 | 0.9933 | 0.4967 | 0.8863 |
| Fe4 | 0.1579 | 0.8401 | 0.4473 | Fe1 | 0.1588 | 0.8412 | 0.4472 |
| Fe5 | 0.1687 | 0.8237 | 0.2206 | Fe2 | 0.1744 | 0.8256 | 0.2216 |
| Fe6 | 0.1579 | 0.3175 | 0.4473 | Fe3 | 0.1588 | 0.3177 | 0.4472 |
| Fe7 | 0.1688 | 0.3465 | 0.2207 | Fe4 | 0.1744 | 0.3488 | 0.2216 |
| Fe8 | 0.6802 | 0.8400 | 0.4475 | Fe5 | 0.6823 | 0.8412 | 0.4472 |
| Fe9 | 0.6468 | 0.8242 | 0.2217 | Fe6 | 0.6512 | 0.8256 | 0.2216 |
| Fe10 | 0.8279 | 0.1736 | 0.7804 | Fe7 | 0.8256 | 0.1744 | 0.7784 |
| Fe11 | 0.8399 | 0.1584 | 0.5531 | Fe8 | 0.8412 | 0.1588 | 0.5529 |
| Fe12 | 0.8278 | 0.6508 | 0.7790 | Fe9 | 0.8256 | 0.6512 | 0.7784 |
| Fe13 | 0.8401 | 0.6808 | 0.5530 | Fe10 | 0.8412 | 0.6824 | 0.5529 |
| Fe14 | 0.3506 | 0.1737 | 0.7789 | Fe11 | 0.3488 | 0.1744 | 0.7784 |
| Fe15 | 0.3179 | 0.1584 | 0.5529 | Fe12 | 0.3176 | 0.1588 | 0.5529 |
| Sn1 | 0.0011 | 0.0015 | 0.1054 | Sn1 | 0.0000 | 1.0000 | 0.1057 |
| Sn2 | 0.0032 | 0.0001 | 0.8939 | Sn2 | 0.0000 | 0.0000 | 0.8943 |
| Sn3 | 0.6655 | 0.3325 | 0.4388 | Sn3 | 0.6667 | 0.3333 | 0.4391 |
| Sn4 | 0.6629 | 0.3319 | 0.2282 | Sn4 | 0.6667 | 0.3333 | 0.2296 |
| Sn5 | 0.3346 | 0.6665 | 0.7717 | Sn5 | 0.3333 | 0.6667 | 0.7704 |
| Sn6 | 0.3325 | 0.6657 | 0.5616 | Sn6 | 0.3333 | 0.6667 | 0.5611 |
| Sn7 | -0.0024 | -0.0010 | 0.3312 | Sn7 | 1.0000 | 1.0000 | 0.3313 |
| Sn8 | 0.0002 | -0.0006 | 0.6691 | Sn8 | 1.0000 | 1.0000 | 0.6688 |
| Sn9 | 0.6678 | 0.3326 | 0.6634 | Sn9 | 0.6667 | 0.3333 | 0.6627 |
| Sn10 | 0.6689 | 0.3356 | 0.0023 | Sn10 | 0.6667 | 0.3333 | 0.0019 |
| Sn11 | 0.3359 | 0.6689 | 0.9970 | Sn11 | 0.3334 | 0.6667 | 0.9980 |
| Sn12 | 0.3299 | 0.6653 | 0.3370 | Sn12 | 0.3334 | 0.6667 | 0.3373 |

**Table S3.** An example configuration of Fe and Mn atoms in the relaxed Fe$_{3-x}$Co$_x$Sn$_2$ unit cell for Co substitution in Site 2 (Table S1), as determined by DFT calculations. The positions are given as fractional coordinates.

| Fe$_{2.5}$Co$_{0.5}$Sn$_2$ | | | | Fe$_2$CoSn$_2$ | | | |
|---|---|---|---|---|---|---|---|
| Atom | x | y | z | Atom | x | y | z |
| Co1 | 0.4970 | 0.5048 | 0.1102 | Co1 | 0.4947 | 0.5053 | 0.1098 |
| Co2 | 0.5072 | 0.5025 | 0.8906 | Co2 | 0.5053 | 0.4947 | 0.8902 |
| Co3 | 0.4966 | 0.9976 | 0.1104 | Co3 | 0.4947 | 0.9893 | 0.1098 |
| Fe1 | 0.5145 | 0.0210 | 0.8865 | Co4 | 0.5053 | 0.0107 | 0.8902 |
| Fe2 | 0.0291 | 0.5174 | 0.1161 | Co5 | 0.0107 | 0.5053 | 0.1098 |
| Fe3 | 0.9876 | 0.4940 | 0.8862 | Co6 | 0.9893 | 0.4947 | 0.8902 |
| Fe4 | 0.1566 | 0.8388 | 0.4467 | Fe1 | 0.1597 | 0.8403 | 0.4464 |
| Fe5 | 0.1597 | 0.8212 | 0.2192 | Fe2 | 0.1742 | 0.8258 | 0.2155 |
| Fe6 | 0.1570 | 0.3159 | 0.4468 | Fe3 | 0.1597 | 0.3193 | 0.4464 |
| Fe7 | 0.1590 | 0.3421 | 0.2192 | Fe4 | 0.1742 | 0.3484 | 0.2155 |
| Fe8 | 0.6767 | 0.8370 | 0.4469 | Fe5 | 0.6807 | 0.8403 | 0.4464 |
| Fe9 | 0.6362 | 0.8206 | 0.2164 | Fe6 | 0.6516 | 0.8258 | 0.2155 |
| Fe10 | 0.8348 | 0.1745 | 0.7808 | Fe7 | 0.8258 | 0.1742 | 0.7845 |
| Fe11 | 0.8393 | 0.1561 | 0.5533 | Fe8 | 0.8403 | 0.1597 | 0.5536 |
| Fe12 | 0.8373 | 0.6518 | 0.7820 | Fe9 | 0.8258 | 0.6516 | 0.7845 |
| Fe13 | 0.8384 | 0.6786 | 0.5532 | Fe10 | 0.8403 | 0.6807 | 0.5536 |
| Fe14 | 0.3558 | 0.1713 | 0.7822 | Fe11 | 0.3484 | 0.1742 | 0.7845 |
| Fe15 | 0.3185 | 0.1579 | 0.5530 | Fe12 | 0.3193 | 0.1597 | 0.5536 |
| Sn1 | -0.0004 | 0.0028 | 0.1041 | Sn1 | 0.0000 | 0.0000 | 0.1011 |
| Sn2 | 0.0055 | 0.0039 | 0.8962 | Sn2 | 0.0000 | 0.0000 | 0.8989 |
| Sn3 | 0.6643 | 0.3308 | 0.4383 | Sn3 | 0.6667 | 0.3333 | 0.4374 |
| Sn4 | 0.6497 | 0.3273 | 0.2243 | Sn4 | 0.6667 | 0.3333 | 0.2211 |
| Sn5 | 0.3451 | 0.6639 | 0.7736 | Sn5 | 0.3333 | 0.6667 | 0.7789 |
| Sn6 | 0.3313 | 0.6642 | 0.5618 | Sn6 | 0.3333 | 0.6667 | 0.5626 |
| Sn7 | -0.0088 | -0.0042 | 0.3298 | Sn7 | 0.0000 | 0.0000 | 0.3281 |
| Sn8 | 0.0028 | -0.0019 | 0.6701 | Sn8 | 0.0000 | 0.0000 | 0.6719 |
| Sn9 | 0.6715 | 0.3323 | 0.6647 | Sn9 | 0.6667 | 0.3333 | 0.6666 |
| Sn10 | 0.6719 | 0.3405 | 0.0035 | Sn10 | 0.6667 | 0.3333 | 0.0019 |
| Sn11 | 0.3421 | 0.6749 | 0.9986 | Sn11 | 0.3333 | 0.6667 | 0.9981 |
| Sn12 | 0.3238 | 0.6627 | 0.3352 | Sn12 | 0.3333 | 0.6667 | 0.3334 |

**Table S4.** Average Fe-Sn and Sn-Sn bond lengths in the optimized structures of $Fe_{3-x}M_xSn_2$, demonstrating that the more pronounced changed of the unit cell along the $c$ axis is caused mainly by changes in the interlayer Fe–Sn distances.

|  | In-Plane Distances (Å) | | Interlayer Distances (Å) | |
| --- | --- | --- | --- | --- |
|  | $d$(Sn–Sn) honeycomb | $d$(Fe–Sn) kagomé | $d$(Fe–Sn) kagomé-kagomé | $d$(Fe–Sn) kagomé-honeycomb |
| $Fe_2MnSn_2$ | 3.093 | 2.682 | 2.794 | 2.712, 2.731 |
| $Fe_{2.5}Mn_{0.5}Sn_2$ | 3.082 | 2.673 | 2.789 | 2.716, 2.738 |
| $Fe_3Sn_2$ Calcd. | 3.077 | 2.671 | 2.783 | 2.721, 2.735 |
| (*Exp.*) | (*3.086*) | (*2.679*) | (*2.799*) | (*2.727, 2.731*) |
| $Fe_{2.5}Co_{0.5}Sn_2$ | 3.077 | 2.676 | 2.773 | 2.724, 2.754 |
| $Fe_2CoSn_2$ | 3.073 | 2.666 | 2.767 | 2.728, 2.732 |